\documentclass[a4paper,10pt]{article}

\usepackage{graphicx}

\title{\textbf{Supernovae: Explosions in the Cosmos}}
\author{\textbf{ $\hbox{P K Suresh}^{\dagger}$ and $\hbox{V H Satheesh Kumar}^{\star}$}\\ School of Physics, University of Hyderabad, HYDERABAD 500 046\\ $^\dagger$ \textit{pkssp@uohyd.ernet.in},  $^\star$ \textit{vhsatheeshkumar@yahoo.co.uk}}

\begin{document}

\maketitle

\begin{abstract}
In this article, a broad perspective of supernovae, their classification and mechanism  is given. Later, the astrophysical significance of supernovae is discussed in brief.
\end{abstract}

\section{Intoduction}
Supernova is one of the most violent phenomena that occur in the universe. Supernovae are really bright -- about 10 billion times as luminous as the Sun. They tend to fade over months or years. The energy output of a supernova surpasses that of the galaxy as a whole! When one such occurs in our own Galaxy, that can be seen even in day light for a few weeks. 

From ancient times astronomers have recorded the appearance of novae-new stars- in the sky at position where nothing was seen before. But later, it has been proved that this is the destruction of a star in a glorious explosion rather the appearance of a ``guest star". The brightest supernova, perhaps, in recorded human history, appeared in the planet Earth's sky in the year 1006 AD, according to the records from the Far East. The remnant is still visible to modern astronomers in the southerly constellation of Lupus. Another supernova event was recorded by the Chinese chronicler Ming Taun-Lin in the year 1054, which is seen today as Crab Nebula. One more such was observed by Tycho Brahe in 1572, now known as Tycho's Supernova Remnant. The last supernova to be seen in our Galaxy, the Milky Way, was in 1604 by the famous astronomer Kepler. These are the four catalogued historical supernovae. The brightest since then was supernova 1987A which happend on $24^{th}$ February 1987 in the Large Magellanic Cloud, a small satellite galaxy of our Milky Way.

\begin{figure}
   \centering
   \includegraphics[scale=0.53]{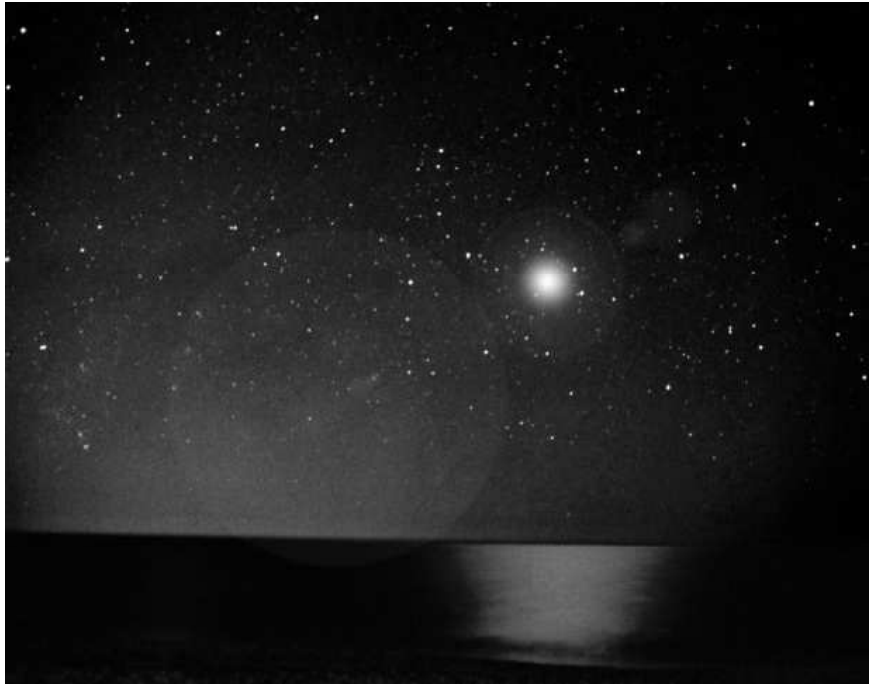}
   \caption{Artist concept of SN1006. Credit and Copyright:Tun\c{c} Tezel.}
   \label{sn1006}
   \end{figure}
   
\begin{figure}
   \centering
   \includegraphics[scale=1.35]{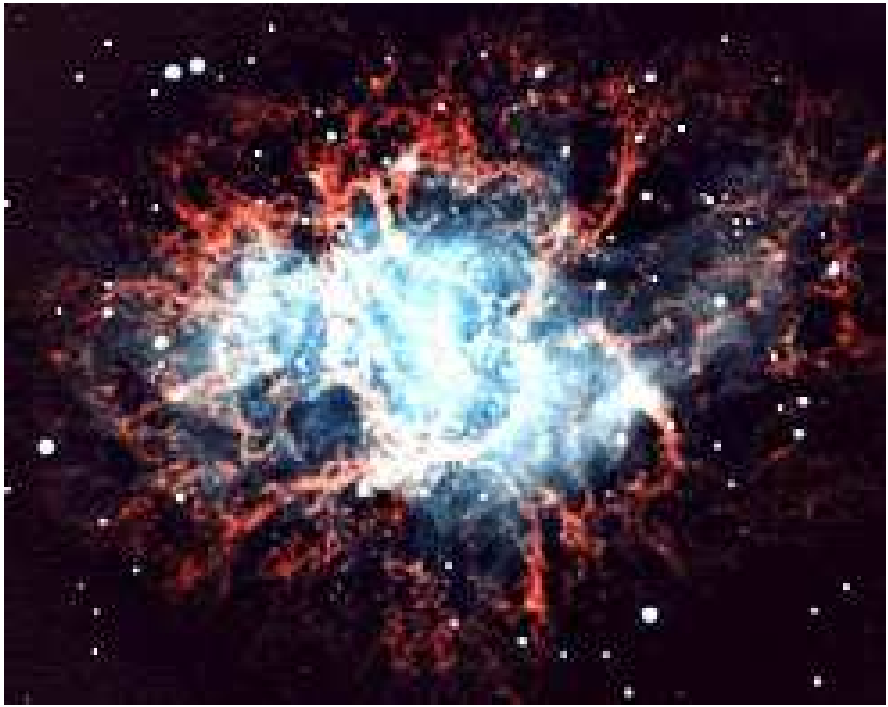}
\caption{Crab nebula, the remnant of SN1054. Photo from Hale 5m Telescope. \copyright \space   Caltech/Pasachoff/Milan.}
   \label{crab}
\end{figure}
  
\begin{figure}
   \centering
   \includegraphics[scale=1.2]{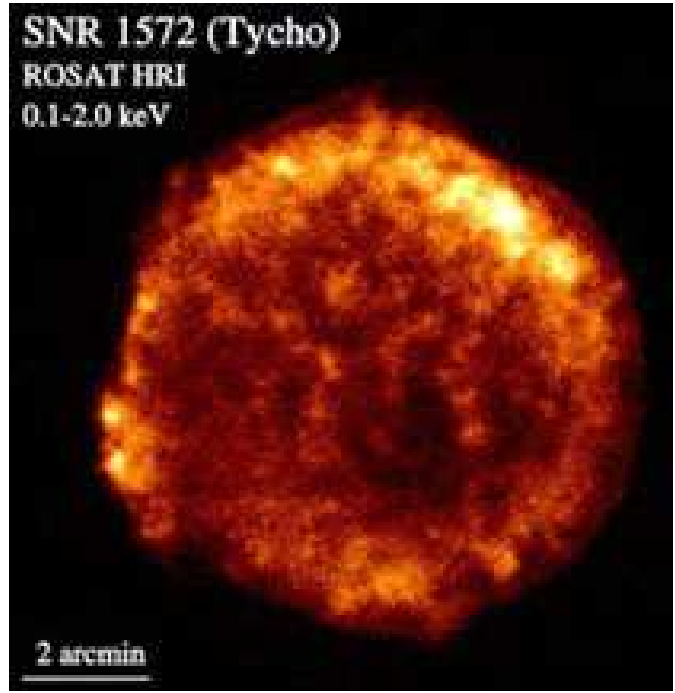}
   \caption{Tycho Supernova remnant. Credit: S.L.Snowden (NASA/GSFC/ LHEA).}
   \label{tycho}
\end{figure}

\begin{figure}
   \centering
   \includegraphics[scale=1.1]{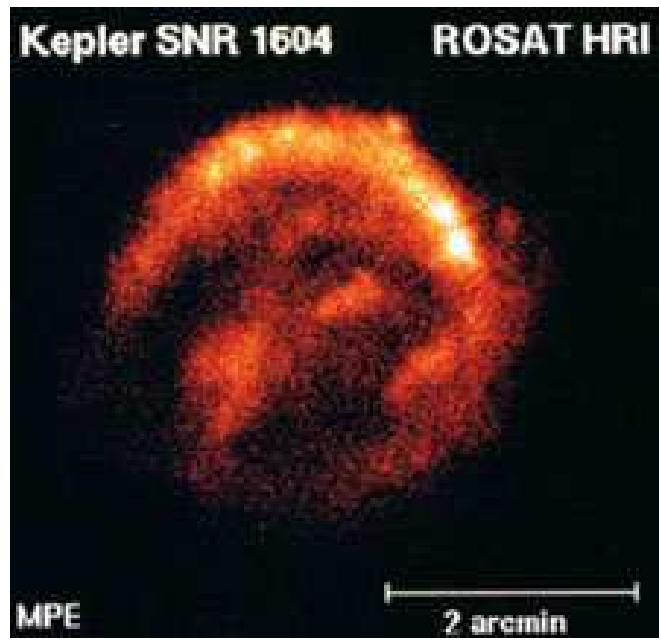}
   \caption{Remnant of SN 1604. Image credit: Max-Planck-Institut für extraterrestrische Physik (MPE).}
   \label{kepler}
\end{figure}

\begin{figure}
   \centering
   \includegraphics[scale=1]{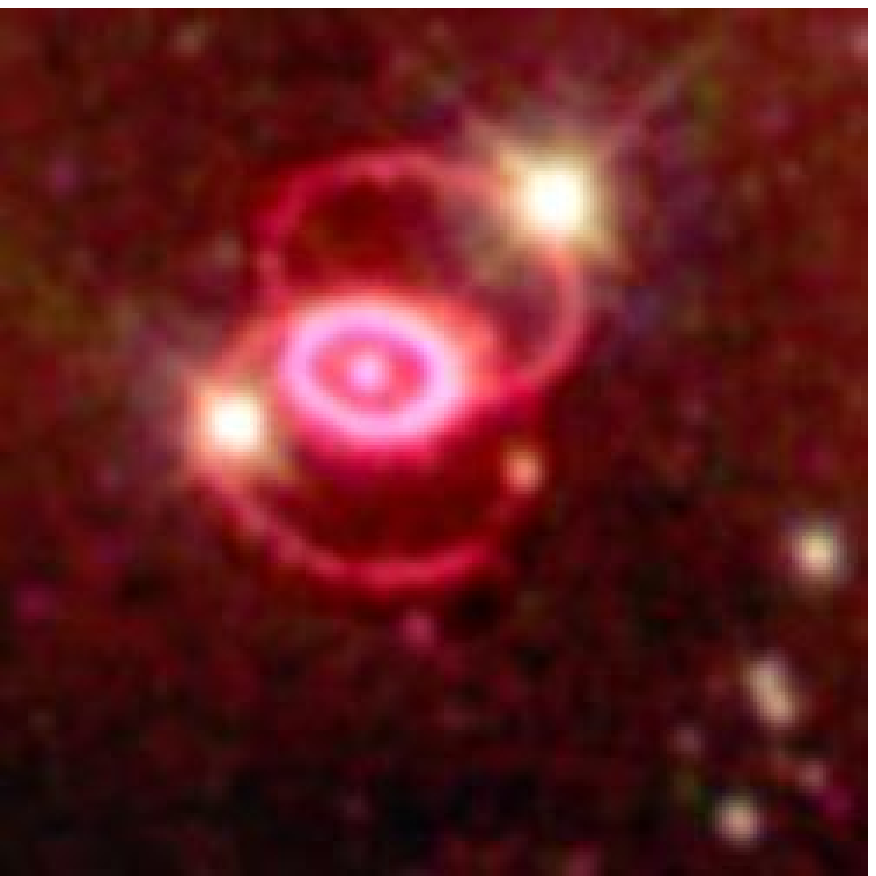}
\caption{SN 1987A in Large Magellanic Cloud. The image was taken by the Hubble Space Telescope (HST) in 1999. (NASA/STScI).}
   \label{sn87a}
\end{figure}

Modern study of supernovae started in 1934 with Wlter Baade and Fritz Zwiky and today the two main groups searching for supernovae are ``Supernova Cosmology Project" led by Saul Perlmutter of Lawrence Berkley Laboratory and ``High-Z Supernova Search" led by Brian Schmidt of the Mt.Stromlo and Siding Observatories. As of 2005 January 1, 3049 supernovae have been discovered since supernovae first really began to be catalogued in 1885. The supernova designation, in the standard form, consisting of the `SN ' prefix followed by the year it is discovered, and then (if there were multiple supernovae in a particular year) either the upper-case letters `A' through `Z' (for the first 26 SNe discovered in a given year) or the lower-case letters `aa', `ab', etc. (for the 27th, 28th, etc., SNe discovered in a given year).

Catching a supernova at its peak proves splitting hair as one cannot predict their occurance and schedule the observing time at the world's largest telescopes in advance. More over they are fleeting and must be observed carefully multiple times within the first few weeks, if they are discovered after passing the peak brightness, that will be of little use as the peak brightness is essential for calibration.

\section{Classification}

Supernovae come in two main observational varieties. The classification is generally done on the optical spectra (taken near their maximum) but, to some extent, also their light curves. The first two main classes of SNe were identified by Minkowski in 1941 based on the presence or absence of hydrogen lines in their spectra. Those whose optical spectra exhibit hydrogen lines are classified as Type II, while hydrogen-deficient SNe are designated Type I.

SNe I are further subdivided according to the detailed appearance of their early-time spectrum. SNe Ia are charecterised by strong absorption near 6150 \AA{} corresponding to SiII, example is SN1996X. SNe Ib lack this feature but instead show prominent HeI lines, e.g. SN1999dn. SNe Ic such as SN1994I, have neither the SiII nor the HeI lines. In many cases there is little distinction between the later two types and they are designated as SN Ib/c.

Four subclasses of SN II are commonly mentioned in literature, viz IIP, IIL, IIb and IIn. SNIIP (plateau) and SNIIL (Linear) constitute the bulk of all SNII and are often referred as normal SNII. Celebrated SN1987A belongs to this class. The subclssification is made according to the shape of optical light curves. The luminosity of SNIIP stops declining shortly after maximum forming plateau 2-3 months long. SNIIL, on the other hand, show a linear, uninterrupted luminosity decline. Indeed there is no clear spectral differences between these two types, but  their progenitors do differ by the amount of H they have in their envelope. A few objects have been found to have early time spectra similar to Type II and late time spectra similar to Type Ib/c. For this reason they have been called SN IIb. The first such was SN1987M. A number of peculiar SNII have been grouped into the class of SNIIn, where stands for `narrow emission lines'. The spectra of these objects have a slow evolution and are dominated by strong Balmer emission lines.

Most of the time, SN Ib/c and SNIIn are associated with Gamma Ray Bursts. To denote these particularly energetic objects, the term "hypernova" has been used, although its meaning is not well defined or universally accepted.

\begin{figure}
   \centering
   \includegraphics[scale=0.45]{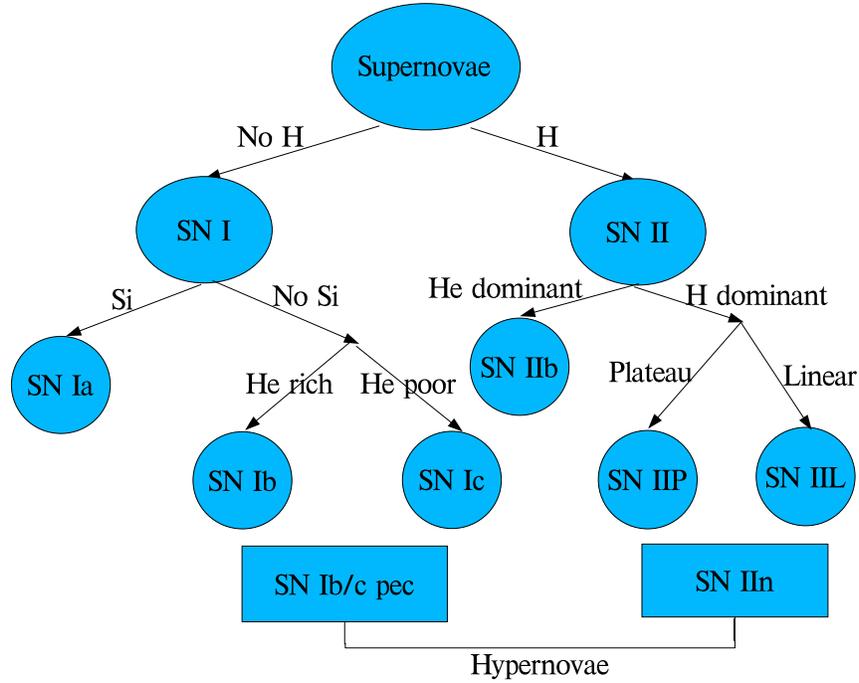}
   \caption{The current classification scheme of supernovae.}
   \label{taxonomy}
\end{figure}

\section{Mechanism}

Physically, there are two fundamental types of supernovae,based on what mechanism powers them: the thermonuclear SNe and the core-collapse ones. Only SNe Ia are thermonuclear type the rest are formed by core-collapse of a massive star. SNe Ia are discovered in all types of galaxies and are not associated with the arms of spirals as strongly as other SN types. The other type SNe mostly appear in spiral galaxies and have been associated with parent population of massive stars.

\subsection{Thermonuclear Supernovae}
 
\begin{figure}
   \centering
   \includegraphics[scale=0.85]{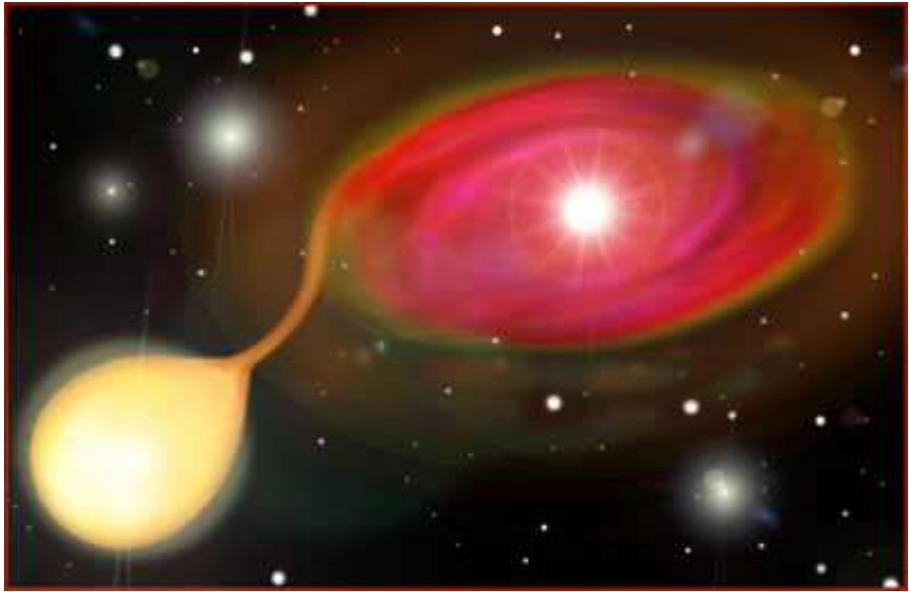}  
   \caption{A computer generated image showing White Dwarf accreting matter from companion Red Giant. Source: Lawrence Berkeley National Laboratory, University of California, Berkeley.}
   \label{supernova I}
\end{figure}

The most  accepted theory of SNe Ia is that they are the result of a white dwarf accreting matter from a nearby companion star, typically a red giant, until it reaches the Chandrasekhar limit. If a white dwarf is in a close binary system with a main sequence star, when the main sequence star expands into a giant or supergiant, it will start to dump gas onto the white dwarf. When the mass of the white dwarf is nudged up to the Chandrasekhar limit, it is no longer stable against collapse. Then the radius decreases resulting in increase of density and temperature. At the new higher density and temperature, the fusion of carbon and oxygen into iron occurs in a runaway fashion. The white dwarf is converted into a fusion bomb, and is blown off completely by the explosion without leaving any remnant behind. The amount of energy released in the explosion is about $10^{44}$ joules, as much energy as the Sun has radiated away during its entire lifetime. 

The spectrum of a type Ia supernova contains no hydrogen or helium lines because the white dwarf that is blown apart consists of carbon and oxygen.The spectrum of a type Ia supernova contains silicon lines because silicon is one of the products of fusing carbon and oxygen.

\subsection{Core Collapse Supernovae}
  
\begin{figure}
   \centering
   \includegraphics[scale=1.1]{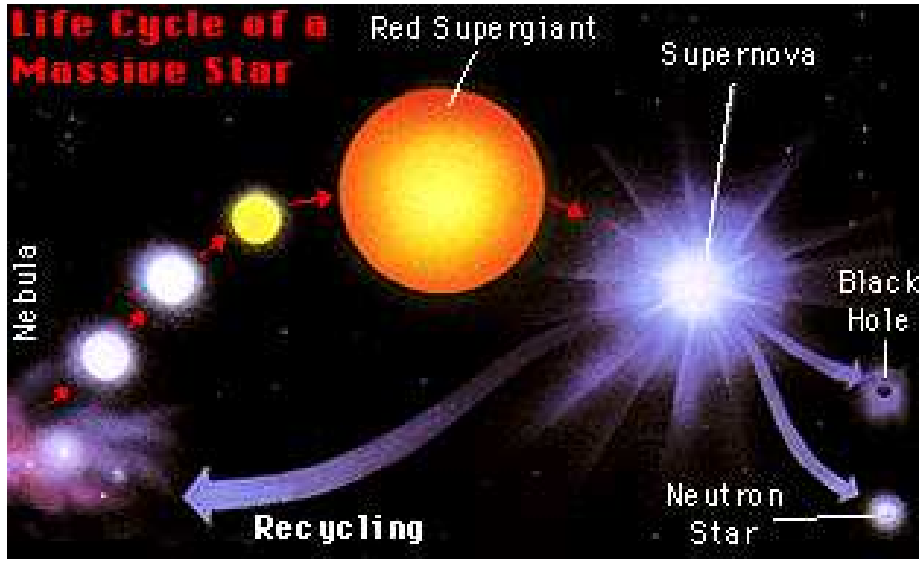}
   \caption{Life cycle of a massive star. A massive star ends up as Neutron star or Black Hole after the supernova explosion. Source NASA's ``Imagine the Universe!" }
   \label{supernova II}
\end{figure}
 
As massive red supergiants become old, they produce ``onion layers" of heavier and heavier elements in their interiors. However, stars will not fuse elements heavier than iron. Fusing iron doesn't release energy, instead it uses up energy! Thus, as no more nuclear fusion is possible, a core of iron builds up in the centers of massive supergiants. Eventually, the iron core reaches something called the Chandrasekhar Mass, which is about 1.4 times the mass of the Sun. When something is this massive, not even electron degeneracy pressure can hold it up, so the core collapses! Then, two important things happen, one is the protons and electrons are pushed together to form neutrons and neutrinos. Even though neutrinos don't interact easily with matter, at densities as high as they are here, they exert a tremendous outward pressure. The other is, outer layers fall inward when the iron core collapses. When the core stops collapsing (this happens when the neutrons start getting packed too tightly -- neutron degeneracy), the outer layers crash into the core and rebound, sending shock waves outward. These two effects -- neutrino outburst and rebound shock wave -- cause the entire star outside the core to be blow apart in a huge explosion, that is what is called, type II supernova!
 
The collapsed core is also left behind by a type II supernova explosion. If the mass of the core is less than 2 or 3 solar masses, it becomes a neutron star. If more than 2 or 3 solar masses remains, not even neutron degeneracy pressure can hold the object up, and it collapses into a black hole. 
 
\section{Astronomical Significance}

The supernovae are very interesting events as they help us in understanding some astrophysical issues such as stellar evolution, stellar mass loss, collapse and explosion physics, radiative hydrodynamics, galactic structure and many more. They trigger new star formation in the galaxy, which are heavy element rich.

The fundamental aim of cosmology (the science of study of the Universe as a whole) is to test its theoretical models against observations. Today, all most all the cosmological observations come from cosmic microwave background and supernovae. Supernovae are the tools at hand(!) for measuring the universe, that is, its mass density, its large scale curvature and its fate. Type Ia SNe have become very popular in the last decade because of their role in determining the geometry of the Universe with their high luminosity and relatively small luminosity dispersion at maximum. These results have surprised everybody by telling that `the universe is accelerating' and giving new turns to the idea of dark energy and dark matter. 

Recently there has been a revived interest in the concept of extra dimensions, which is built upon the idea that our universe ( referred as ``brane") is a part of the higher dimension spacetime called ``bulk". In this scenario all the standard model particles, i.e. all the matter that we know of, is confined to ``brane" and only gravity can escape to higher dimensions of the universe. To test this idea, partical physicists all over the world are planning for an accelerator experiment at CERN, in near future. But the same idea can also be tested by studying the cooling rate of supernova ( the authors are presently working in this area). This is a novel approach to extra-dimensions using astronomical observations. That is why supernovae have attracted the attentions of many people working in scattered branches of physics and astronomy. After all, we did not have to build anything nearly as formidable as multi-billion dollar giant accelerators and detectors for particle physics!!

During the supernova, a tremendous amount of energy is released. Some of that energy is used to fuse elements even heavier than iron! This is where such heavy elements like gold, silver, zinc and uranium come from! The material that gets ejected into space as a result of the supernova becomes part of the interstellar medium(ISM). New stars and planets form from this interstellar medium. Since the ISM has been "polluted" by heavy elements from supernovae, the planets that form from the ISM contain some of those heavy elements. That is how you and me have many heavy elements in our body. You must have realized by now, that ``we are all made of star stuff!!''\\[-0.8cm]

\section{References}
\bibliography{supernova}
1. H Bethe \textit{Supernovae}, Physics Today, September 1990 \\
2. M Turatto  \textit{Classification of Supernovae}, astro-ph/0301107\\
3. S Perlmutter \textit{Supernovae, Dark Energy and the Accelerating Universe}, Physics Today, April 2003\\
4. Supernova Cosmology Project, \textit{http://www-supernova.lbl.gov/}\\
5. High-Z Supernova Search, \textit{cfa-www.harvard.edu/cfa/ oir/Research/supernova/HighZ.html}\\
6. V Barger, T Han, C.Kao and R.-J.Zang \textit{Astrophysical Constraints on Large Extra Dimensions}, hep-ph/9905474

\end{document}